\documentclass[11pt,a4paper]{article}
\usepackage[table]{xcolor}
\usepackage[hyperref]{naaclhlt2018}
\usepackage{times}
\usepackage{latexsym}
\usepackage{graphicx}
\usepackage{url}
\usepackage{amsmath}
\usepackage{amsfonts}
\usepackage{ amssymb }
\usepackage{booktabs}
\usepackage{tabularx}
\usepackage{multirow}
\usepackage{collcell}
\usepackage{hhline}
\usepackage{pgf}
\usepackage{diagbox}
\newcommand\gray{gray}

\newcommand\ColCell[1]{%
\pgfmathparse{#1/112<.4?1:0}
\ifnum\pgfmathresult=0\relax\color{white}\fi%
\pgfmathparse{1-#1/112}%
\expandafter\cellcolor\expandafter[%
\expandafter\gray\expandafter]\expandafter{\pgfmathresult}#1}

\newcolumntype{E}{>{\collectcell\ColCell}c<{\endcollectcell}}
\aclfinalcopy % Uncomment this line for the final submission
\title{Multiple Document Representations from News Alerts for Automated Bio-surveillance Event Detection} %Classification}
 \author{
	Aaron Tuor  \and Fnu Anubhav \and Lauren Charles \\ 
         Pacific Northwest National Laboratory \\ Richland, WA}

\date{}

\begin{document}
\maketitle

%==========================================================
%====================	ABSTRACT=============================
%==========================================================
\begin{abstract}
Due to globalization, geographic boundaries
no longer serve as effective shields for the
spread of infectious diseases. In order to aid
bio-surveillance analysts in disease tracking,
recent research has been devoted to developing information retrieval and analysis methods utilizing the vast corpora of publicly available documents on the internet. In this work,
we present methods for the automated retrieval
and classification of documents related to active public health events. We demonstrate classification performance on an auto-generated
corpus, using recurrent neural network, TF-IDF, and Naive Bayes log count ratio document representations. By jointly modeling the
title and description of a document, we achieve
97\% recall and 93.3\% accuracy with our best
performing bio-surveillance event classification model: logistic regression on the combined output from a pair of bidirectional recurrent neural networks.
\end{abstract}
%==========================================================
%====================	INTRODUCTION=========================
%==========================================================
\section{Introduction} \label{sec:intro}
Automated mining and analysis of open source
web-based content provides a powerful aid in
effective monitoring of public health concerns
\cite{hartley2013overview, volkova2017forecasting, pavalanathan2016discourse, milinovich2014internet, charles2015using}. 
A deluge of potentially relevant content can quickly be generated
from web monitoring services; overwhelming the
cognitive capabilities of human analysts tasked
with identifying and assessing current biosurveillance events. This glut of information combined
with the great cost associated with missing timely
events of importance affirms the need for reliable
downstream classification of retrieved documents.

To train a biosurveillance current event document classification system, we assembled an analyst labeled data set of 30,893 articles from Google
Alerts\footnote{https://www.google.com/alerts} \cite{kataru2016systems}. The news alerts provide
multiple views of a news article: a title and description. These multiple views are a distinctive
feature of our data set; there are two natural language texts, a title and description, associated with
each news article. This offers a direct avenue to
compare the effectiveness of joint document representations for traditional bag-of-N-grams models against their more sophisticated neural network
successors, which, to our knowledge, is a novel
contribution. In this work, we explore joint models of multiple document representations for classification. We compare the effectiveness of joint
modeling for both state-of-the-art Recurrent Neural Networks (RNNs) and traditional bag-of-N-grams models. We find that classification performance is enhanced more significantly by the joint
modeling of multiple text representations than by
a choice of document model between bag-of-N-grams models and RNN.

\section{Related Work}
Recurrent neural networks with LSTM \cite{hochreiter1997long} cells are often fa-
vored as document models for their potential to
model long term word dependencies. \citet{miyato2016adversarial}  use bidirectional \cite{schuster1997bidirectional} LSTMs with unsupervised pre-training and
adversarial noise regularization on the initial embedding vectors. They concatenate the final hidden states of the forward and backward LSTM
as the document representation for classification.
\citet{liu2015fine} use max pooling over the hidden
states of a bidirectional LSTM for classification.
In contrast to these works, we use average pooling
over the hidden states concatenated with the final
hidden state to represent the text as a vector.

Joint modeling using multiple representations
has demonstrated reliable performance gains for
several neural network document classification
models. \citet{limsopatham2016modelling} create
separate document models from two sets of pre-trained word vectors using texts from the classification target domain and a large unlabeled corpus. 
They found vectorizing the word embedding matrices with separate convolutional networks performed better than using combined matrices as input to a single convolutional network.
\citet{mccann2017learned} use GloVe \cite{pennington2014glove} word embeddings in tandem with
pre-trained word embeddings from the English-to-German translation task. \citet{johnson2016supervised} improve performance using joint bidirec-
tional LSTM representations of a document; one
network has weights fixed after unsupervised pre-training while the other is learned from labeled
documents. They showed similar gains using this
same semi-supervised set-up but with convolutional networks in place of LSTMs.

\begin{table*}
	\begin{tabularx}{\linewidth}{l  c  c  c  c  }
		{\bf Corpus}&{\bf Number of Texts}&{\bf Vocabulary Size}&{\bf Median length} &{\bf Average Length}\\
		\midrule
		Unlabeled-Title&1,013,804 &103,214 &12 & 13 \\
		\cellcolor[gray]{0.9}Unlabeled-Description&\cellcolor[gray]{0.9}1,149,959&\cellcolor[gray]{0.9}241,060 & \cellcolor[gray]{0.9}30& \cellcolor[gray]{0.9}47 \\
		Labeled-Title&30,893 & 64,886&13 &14  \\
		\cellcolor[gray]{0.9}Labeled-Description&\cellcolor[gray]{0.9}30,893 &\cellcolor[gray]{0.9}24,715 &\cellcolor[gray]{0.9}37 &\cellcolor[gray]{0.9}  82\\
	\end{tabularx}
	\caption{Statistics for assembled labeled and unlabeled corpora. } \label{tab:dataset}
\end{table*}

%==========================================================
%====================	METHODS=========================
%==========================================================
\section{Methods} 
	In this section we describe our methods for bio-surveillance text retrieval and classification.
%==========================================================
%====================	DOCUMENT RETRIEVAL=========================
%==========================================================
	\subsection{Retrieval} 
	We retrieve potentially relevant documents con-
taining disease names, acronyms, and synonyms
through Google Alerts service. The set of queries,
created by biosurveillance analysts, return alerts
sent to a dedicated electronic mailbox, which is
periodically mined for new content. An example
alert returned for the ebola query:
\begin{description}
\itemsep0em 
\small
\item[title] Ebola survivors sue state over disappearance of aid money
\item[description] Two Ebola survivors are to sue the government of Sierra Leone in the first international court case aimed at throwing light on what happened to some of the millions of dollars siphoned off from funds intended to help fight the disease. The case, filed...
\end{description}
        The alerts provide a title, description, and URL link to the full text. 
        Table \ref{tab:dataset} records the data-set statistics. 30,893 title/description document pairs were hand labeled by analysts,  of which 15,331 were considered events of interest, and 15,362  were considered irrelevant to bio-surveillance.  
       The title/description of 30,893 documents were labeled
at a near 50\% split for the binary classification task
of detecting active biosurveillance events. This
data was labeled by 8 biosurveillance analysts
and reviewed by a lead analyst. About 1M unlabeled alert texts were also retrieved for analysis.
%==========================================================
%====================	DOCUMENT REPRESENTATION=========================
%==========================================================		
	\subsection{Document Representation} \label{sec:doc}	 
	We consider three vector representions of documents, term frequency inverse document frequency ($D_{\texttt{TF}}$), log-count ratio ($D_{\texttt{LCR}}$), and a bidirectional recurrent neural network document representation ($D_{\texttt{RNN}}$). 
	Our vocabulary $V$ is the set of all bigrams and unigrams that appear in the training corpus of $N$ documents. 
	The matrix $\textbf{F} \in \mathbb{R}^{N \times |V|}$ is the document count matrix such that $\textbf{F}_{k,j}$ is the number of times term $j$ occurs in document $k$. 
	The $j$-th element of the $D_{\texttt{TF}}$ vector for document $d_k$ is then:
	\begin{equation}
	D_{\texttt{TF}}(d_k)_j=\frac{\textbf{F}_{k,j}}{\parallel \textbf{F}_{k,:} \parallel_{\infty}} \log{\frac{N}{n_j}}
	\end{equation}
	where $n_j$ is the number of documents in the training corpus where term $j$ appears at least once, 
	and $\parallel * \parallel_{\infty}$ is the max norm of a vector. 
	
	%The log-count ratio is a factor in calculating Naive Bayes classification. 
	We use an add-one smoothed term-count matrix to derive a log-count ratio vector for all terms in the vocabulary. Let $\textbf{p}$ and $\textbf{q}$ be the smoothed positive and negative term count vectors. Then:
	\begin{eqnarray}
	&\textbf{y}_{k}&= \begin{cases} 1 \textrm{ if document } k \textrm{ is an event}\\
	0 \textrm{ otherwise}
	\end{cases}\\
	&\textbf{p} &= \textbf{y}^{\intercal}(\textbf{1} + \textbf{F}) \in \mathbb{R}^{1 \times |V|}\\
	&\textbf{q} &=  (\textbf{\textbf{1} - y})^{\intercal}(\textbf{1} + \textbf{F}) \in \mathbb{R}^{1 \times |V|} \\
	&\textbf{r} &= \log{\frac{\textbf{p} \backslash \parallel \textbf{p} \parallel_1}{\textbf{q} \backslash \parallel \textbf{q} \parallel_1}} \in \mathbb{R}^{1 \times |V|}
	\end{eqnarray}
	where {\bf 1} is a suitably sized vector or matrix of ones, and  $\frac{*}{*}$ is elementwise division. Now our vector representation $D_{\texttt{LCR}}$ is defined as:
	\begin{equation}
	D_{\texttt{LCR}}(d_k) = \texttt{sign}(\textbf{F}_{k,:}) \odot \textbf{r}
	\end{equation}
	where $\odot$ is elementwise multiplication. The \texttt{sign} function binarizes $\textbf{F}$ into a matrix of 1's and 0's.
	
	For the recurrent neural network representation,  $D_{\texttt{RNN}}$, we use the GloVe algorithm to create vectors in $\mathbb{R}^{256}$ representing each word in an enlarged unigram dictionary from the combined corpus of unlabeled documents and a training set of the labeled documents. This allows high quality vector representations of words absent in the training set.
% Preliminary experiments showed that GloVe vectors improve generalization compared to randomly initialized word vectors.  
 \begin{figure}
  \includegraphics[width=\linewidth]{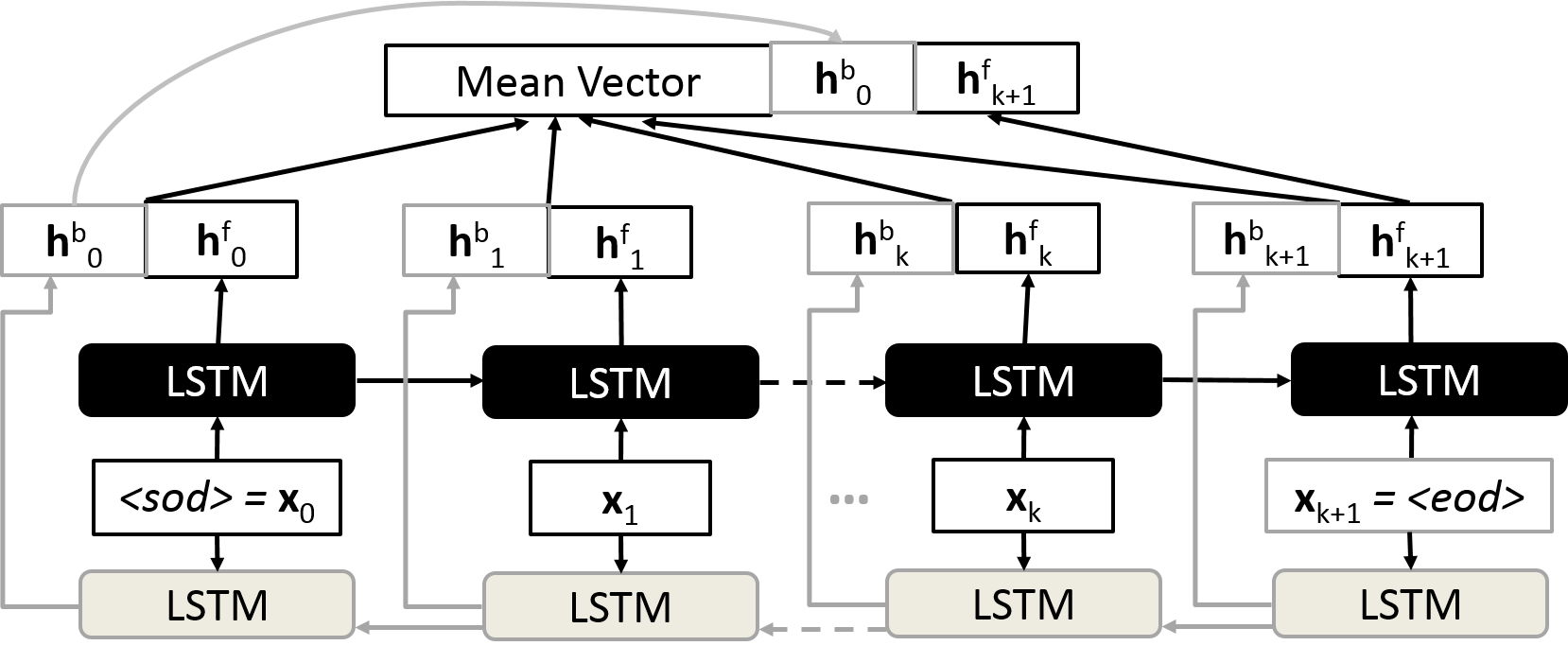}
\caption{Bi-directional LSTM document model} \label{fig:bidir}
\end{figure}
%Following the language model formulation suggested in \cite{schuster1997bidirectional},  

As depicted in figure \ref{fig:bidir}, we model documents from word vector sequences using a bidirectional LSTM. Let $\mathcal{X} = {\bf x}_1, {\bf x}_2, ..., {\bf x}_k$ be the sequence of GloVe vectors associated with a document comprised of $k$ tokens. Let ${\bf h}^f_t$ be the $t$-th hidden state of the forward LSTM and  ${\bf h}^b_t$  the $(k - t)$-th hidden state of the backward LSTM. Consider the hidden states as row vectors, and let   ${\bf h}_t = \begin{bmatrix} {\bf h}^b_t & {\bf h}^f_t \end{bmatrix}$.
Our RNN document vector representation is:
\begin{equation}
D_{\texttt{RNN}} = \begin{bmatrix} \frac{1}{k} * 
\sum_{t=1}^k {\bf h}_t &{\bf h}_{0}^b&{\bf h}_{k+1}^f\end{bmatrix}, 
\end{equation}
the means of the forward and backwards hidden states concatenated with the final LSTM outputs.
\begin{figure*}
\minipage{.33\textwidth}%
\centering
 \includegraphics[width=0.98\linewidth]{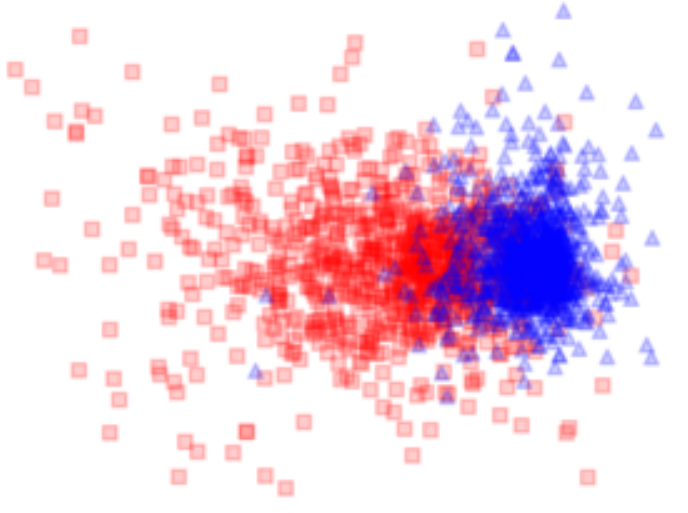}
\caption{Description} \label{fig:desc}
\endminipage
\minipage{.33\textwidth}
\centering
  \includegraphics[width=0.98\linewidth]{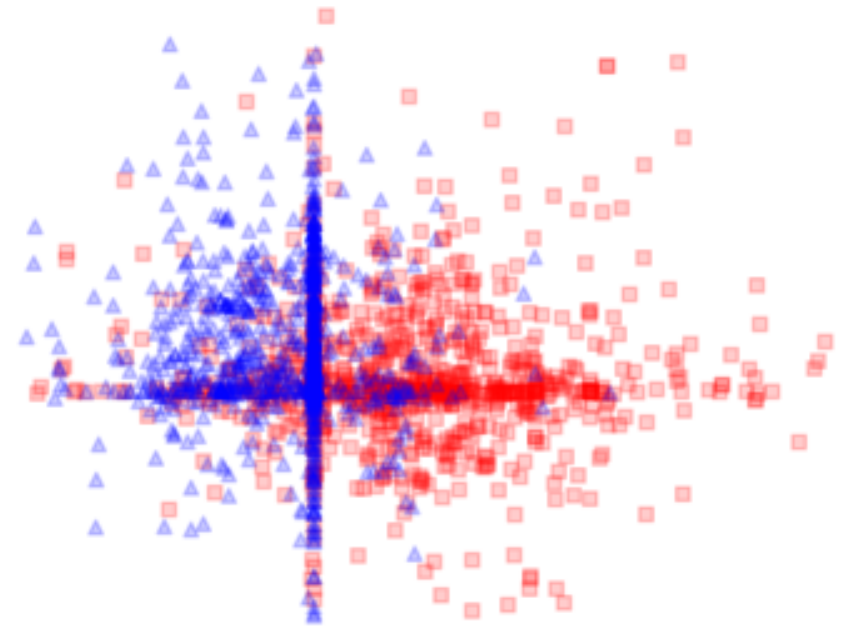}
\caption{Title} \label{fig:title}
\endminipage\hfill
\minipage{.33\textwidth}%
\centering
 \includegraphics[width=0.98\linewidth]{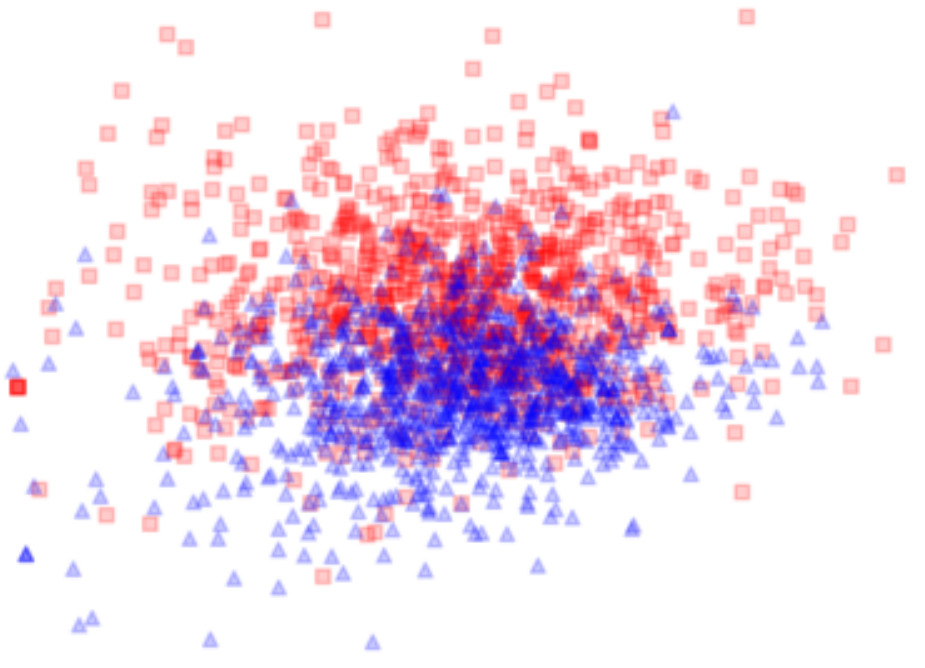}
\caption{Title and Description} \label{fig:both}
\endminipage
\caption{2-D T-SNE projections of RNN document representations.}
\label{fig:vis}
\end{figure*}
\subsection{Classification}
We fit logistic regression models with L2 regularization to classify documents for each of the three document representations $D_{\texttt{TF}}$, $D_{\texttt{LCR}}$, and  $D_{\texttt{RNN}}$. 
For each news article, $d$, in our labeled corpus, we've obtained two views from the Google Alerts; $d_{\texttt{title}}$, the alert title, and  $d_{\texttt{desc}}$, the alert description. For $D_* : * \in \{ \texttt{TF}, \texttt{LCR},  \texttt{RNN}\}$ we consider feature vectors:
\begin{eqnarray} 
&D_{\texttt{*}}\bigl(d_{\texttt{desc}}\bigr), \\
&D_{\texttt{*}}\bigl(d_{\texttt{title}}\bigr) \textrm { and,} \\  
&\begin{bmatrix} D_{\texttt{*}}\bigl(d_{\texttt{desc}}\bigr) & D_{\texttt{*}}\bigl(d_{\texttt{title}}\bigr) \end{bmatrix}
\end{eqnarray}
which are inputs to a logistic regression (LR) classifier using the description, the title, or both the title and description. For the $D_{\texttt{RNN}}$ representation we jointly learn the RNN parameters and logistic regression parameters. For the $D_{\texttt{TF}}$, and $D_{\texttt{LCR}}$ models, separate term count matrices are calculated for the title and description texts.

%==========================================================
%====================	EXPERIMENTS==========================
%==========================================================
\section{Experiments}
Our principal set of experiments is designed to
test the classifier performance across two axes:
the document model, and the text input. We explore nine LR model configurations; three document models (TF, LCR, RNN) each using either text only, description only, or both text and description.  An additional set of experiments assesses performance of the $D_{\texttt{TF}}$ representations for a suite of three other classifiers, Naive Bayes, Support Vector Machine, and Random Forest.  
\subsection{Experiment Set-up}
The joint RNN, logistic regression model is coded in Tensorflow \cite{abadi2016tensorflow}.  
%Optimization is performed using a cross-entropy loss and the Adam \cite{kingma2014adam} method of stochastic gradient descent.
For the $D_{\texttt{LCR}}$ classification model, we adapt open source code\footnote{https://github.com/mesnilgr/nbsvm} from Gr\'egoire Mesnil which implements the logistic regression variant of NB-SVM \cite{wang2012baselines}. 
We implement the $D_{\texttt{TF}}$ classification model for logistic regression and several other classifiers using scikit-learn \cite{scikit-learn}. 

The labeled documents are split into train, de-
velopment, and test sets with 24,715 documents
in the train set, and 3,089 documents each in the
development and test sets. A random search was
performed over hyper-parameters for each model
and results are reported on the test set for the best
performing development set models.

\subsection{Results and Analysis}
\begin{figure}[h]
 \includegraphics[width=.48\textwidth]{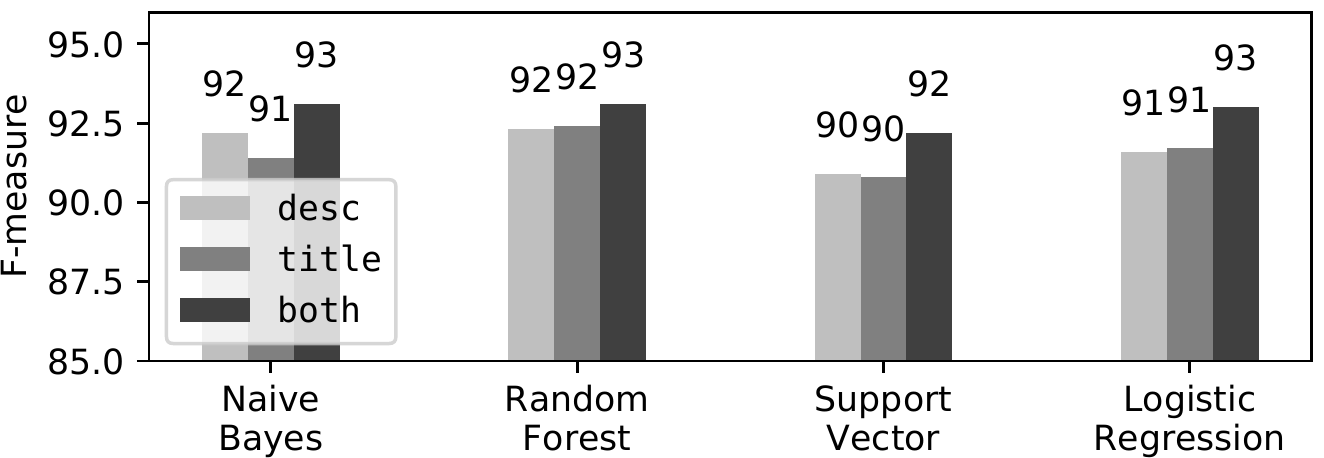}
\caption{F-measure for assorted TF-IDF classifiers} \label{fig:baselines} 
\end{figure}

Table \ref{tab:perf} shows performance for nine logistic regression models. 
Our first observation is that both bag-of-N-grams models and 
the RNN perform better using a joint model of title and description. 
Further, the performance gain from using joint modeling is more significant than the
choice of document vector representation for ac-
curacy and F-measure. Since high recall is critical
for the bio-surveillance application, the joint $D_{\texttt{RNN}}$
representation is most suitable with 97\% recall,
followed by the description $D_{\texttt{RNN}}$ which attains
95.8\% recall. The other two $D_{\texttt{LCR}}$ and $D_{\texttt{TF}}$ joint
representations also give good recall with 95.1\%
and 94.8\% respectively. For the RNN representation, using the longer description text performs
better than using the shorter title text. This speaks
to the RNN representation's capability to model
long term dependencies in the text; a capability
lacking in the simpler bag-of-Ngram models. Figure \ref{fig:baselines} confirms the advantage of a $D_{\texttt{TF}}$ joint representation over a wider range of models.
\begin{table}
	\begin{tabularx}{\linewidth}{l l  c  c  c  c c}
		{\bf Model}&{\bf Text}&{\bf Prec.}&{\bf Rec.}&{\bf F-scr.}&{\bf Acc.} \\
		%=========================logistic regression
		\midrule
		\multirow{3}{*}{$D_{\texttt{TF}}$}&{Desc.}&89.3&94.4&91.8&91.5 \\
		&\cellcolor[gray]{0.95}{Title.}&\cellcolor[gray]{0.95}88.9&\cellcolor[gray]{0.95}94.2&\cellcolor[gray]{0.95}91.5&\cellcolor[gray]{0.95}91.2 \\
		&\cellcolor[gray]{0.85}{Both}&\cellcolor[gray]{0.85}91.0&\cellcolor[gray]{0.85}95.1&\cellcolor[gray]{0.85}93.0&\cellcolor[gray]{0.85}92.8\\
		%===============NBSVM
		\midrule
		&{Desc.}&90.1&93.2&91.6&91.4 \\
		{$D_{\texttt{LCR}}$}&\cellcolor[gray]{0.95}{Title}&\cellcolor[gray]{0.95}91.2&\cellcolor[gray]{0.95}92.3&\cellcolor[gray]{0.95}91.7&\cellcolor[gray]{0.95}91.6 \\
		&\cellcolor[gray]{0.85}{Both}&\cellcolor[gray]{0.85} \textcolor{red}{91.2}&\cellcolor[gray]{0.85}94.8&\cellcolor[gray]{0.85}93.0&\cellcolor[gray]{0.85}92.8 \\
		%============================RNN
		\midrule
		&{Desc.}&89.7&95.8&92.7&92.4 \\
		&\cellcolor[gray]{0.95}{Title}&\cellcolor[gray]{0.95}88.9&\cellcolor[gray]{0.95}94.1&\cellcolor[gray]{0.95}91.4&\cellcolor[gray]{0.95}91.2\\
		\multirow{-3}{*}{$D_{\texttt{RNN}}$}&\cellcolor[gray]{0.85}{Both}&\cellcolor[gray]{0.85}90.4&\cellcolor[gray]{0.85} \textcolor{red}{97.0}&\cellcolor[gray]{0.85} \textcolor{red}{93.6}&\cellcolor[gray]{0.85} \textcolor{red}{93.3} \\
	\end{tabularx}
	\caption{Recall, precision, F-measure, and accuracy for each model using the title and/or description as input. } \label{tab:perf}
\end{table}

Figures  \ref{fig:desc},  \ref{fig:title}, and \ref{fig:both} plot t-SNE \cite{maaten2008visualizing} visualizations of the description, title, and joint RNN representations. The 2D visualization suggests that title and description vectors are mapped onto different manifolds in  high dimensional space. For most models, title and description representations perform similarly, suggesting that performance gains from joint modeling may be from regularizing effects of complementary representations.

\subsection{Online Performance}
In this section, we review online performance of the best performing active public health event classifier (the joint RNN document representation) as deployed in an application monitoring the live stream of online sources. 

The performance results from analyst review of 122 document classification predictions are recorded in Table \ref{tab:review}. High recall is of most importance due to the high cost of false negatives which could lead to overlooking active public health events. The model maintains a high recall with a slight drop when moving from our test set to live data, but shows a significant drop in precision due to a large number of false positives. 

Tables \ref{tab:fneg} and \ref{tab:fpos} show examples of false positives and false negatives from the 122 reviewed online documents, respectively. The false positive examples present text which is similar to text which might describe an active public health event. The false negative examples are clearly about urgent events of interest: a tuberculosis epidemic, a meningitis case, a hepatitis outbreak, and a water source tainted with Legionella. These omissions could be due to lack of coverage for these diseases in the labeled training data or articles which mention past but not current outbreaks of these diseases which would be labeled as non-events by an analyst.

%%%%%%%%%%%%%%%%%%%%%%%%%%%%%%%%%%%%%%%%%%%%%%%%%%%%%%%
% Section 5.1 Analyst Validation
%%%%%%%%%%%%%%%%%%%%%%%%%%%%%%%%%%%%%%%%%%%%%%%%%%%%%%%%%
\begin{table}
\small
\noindent\begin{tabular}{c*{7}{E}}
\diagbox{{\small True}}{\small Pred} & \multicolumn{1}{c}{Event} & \multicolumn{1}{c}{Other}  &\multicolumn{1}{c}{} & \multicolumn{1}{c}{} & \multicolumn{1}{c}{Precision} & \multicolumn{1}{c}{Recall} \\
 Event& 60 & 7   & \multicolumn{1}{c}{} &\multicolumn{1}{c}{} &0.73 &0.90 \\ 
 Other & 22 & 23   & \multicolumn{1}{c}{} &\multicolumn{1}{c}{}  & \multicolumn{1}{c}{} &\multicolumn{1}{c}{}\\
\end{tabular}
\caption{Confusion matrix and metrics from analyst review of classifier.}\label{tab:review}
\end{table} 
\vskip1ex
%%%%%%%%%%%%%%%%%%%%%%%%%%%%%%%%%%%%%%%%%%%%%%%%%%%%%%%
% Section 5.2 False Negative Title Examples
 %%%%%%%%%%%%%%%%%%%%%%%%%%%%%%%%%%%%%%%%%%%%%%%%%%%%%%%%%
 \begin{table}
\small
 \begin{tabular}{|l|}
 \hline
 Tuberculosis epidemic affects the penalties of the Hopper, 
 \\the well and female jail\\\hline
Meningitis\\\hline
Outbreak of Hepatitis A in Sinaloa\\\hline
Legionella found in Rendac water purification plant\\
\hline
 \end{tabular}
 \caption{False negative title examples.}\label{tab:fneg}
 \end{table}
 %%%%%%%%%%%%%%%%%%%%%%%%%%%%%%%%%%%%%%%%%%%%%%%%%%%%%%%
 % Section 5.3 False Positive Title Examples
 %%%%%%%%%%%%%%%%%%%%%%%%%%%%%%%%%%%%%%%%%%%%%%%%%%%%%%%%%
  \begin{table}
 \small
 \begin{tabular}{|l|}
 \hline
 They created an ecological trap to kill the \\
 mosquitoes that produce Zika and Dengue\\\hline
Study: 7\% risk of birth defects in Zika pregnancies\\\hline
Beginning of the end for Ghouta rebels: \\
Thousands flee relentless regime assault\\\hline
PRO/AH/EDR$>$ Salmonellosis, st Agbeni - \\
USA: pet turtles\\
\hline
 \end{tabular}
 \caption{False positive title examples.}\label{tab:fpos}
 \end{table}
%%==========================================================
%====================	CONCLUSION==========================
%==========================================================
\section{Conclusion}
Here, we present simple and effective components
for biosurveillance monitoring. Importantly, the
tools for building a system are openly accessible,
making event detection available to a broad range
of interests. Our best performing method employs
a pair of bidirectional recurrent neural networks,
jointly modeling the text and description of a document with a logistic regression classifier (93.3\%
accuracy; 97\% recall). We find that joint representation models achieve similar performance gains
for both recurrent neural networks and traditional
bag-of-N-grams models.

We are currently exploring a more general class
of joint document representation models which
this work does not address, e.g., combining bag-of-N-grams models with neural network models,
combining log count ratio vectors with TF-IDF
vectors, and other permutations. Finally, we note
that, for other data sets, bag-of-N-grams models may also be enhanced through joint modeling. Traditional feature selection methods \cite{yang1997comparative} can be used to generate multiple document views for joint models through document term thresholding.

\section*{Acknowledgments} 
This work was funded by Department of Homeland Security, Homeland Security Advanced Research Projects Agency-Chemical Biological Division under Agreement \# HSHQPM-17-00077.

\bibliography{event}
\bibliographystyle{acl_natbib}

\end{document}